# Magic Numbers for the Photoelectron Anisotropy in Li-Doped Dimethyl Ether Clusters

*Jonathan V. Barnes, Bruce L. Yoder, and Ruth Signorell\**

ETH Zürich, Laboratory of Physical Chemistry, Vladimir-Prelog-Weg 2, CH-8093, Zürich, Switzerland

\* To whom correspondence should be addressed. E-mail: rsignorell@ethz.ch

**ABSTRACT**

Photoelectron velocity map imaging of Li(CH$_3$OCH$_3$)$_n$ clusters (1 ≤ n ≤ 175) is used to search for magic numbers related to the photoelectron anisotropy. Comparison with density functional calculations reveals magic numbers at n=4, 5, and 6, resulting from the symmetric charge distribution with high s-character of the highest occupied molecular orbital. Since each of these three cluster sizes correspond to the completion of a first coordination shell, they can be considered as "isomeric motifs of the first coordination shell". Differences in the photoelectron anisotropy, the vertical ionization energies and the enthalpies of vaporization between Li(CH$_3$OCH$_3$)$_n$ and Na(CH$_3$OCH$_3$)$_n$ can be rationalized in terms of differences in their solvation shells, atomic ionization energies, polarizabilities, metal-oxygen bonds, ligand-ligand interactions, and by cooperative effects.

## 1. Introduction

Magic numbers play a central role in cluster science (see references on molecular clusters[1-12]). Usually, these magic numbers are related to the high stability of clusters of certain sizes. By contrast, reports on magic numbers related to photoelectron anisotropy are comparatively sparse.[10, 11, 13, 14] This is because measurements of photoelectron angular distributions (PADs) of clusters are not so common and the modelling of cluster PADs is demanding.[5, 10, 11, 13-29] Typically, a prerequisite for the observation of magic numbers in the photoelectron anisotropy is a high cluster symmetry that results in orbitals with high fractional s-character.[10, 13, 14, 24]

In our recent studies,[10, 24] we reported the first observation of magic numbers in the photoelectron anisotropy of solvated electrons in Na-doped clusters of dimethyl ether, ammonia, methanol and water. The studies have revealed that in clusters of high symmetry the solvated electron can delocalize over extended regions, forming symmetric charge distributions of high s-character. However, they have also shown that the direct experimental observation of magic clusters can be hindered by several factors. An important factor is the lack of size selection for the neutral clusters under investigation. This results in PADs that are averages of several cluster sizes, making the detection of magic numbers more difficult. Furthermore, many structural isomers with similar energies can occur in these weakly-bound systems, again making the observation of magic numbers less likely compared with systems that exhibit fewer structural isomers. Our calculations showed that in particular the strong hydrogen bonds in the Na-doped methanol and water clusters result in a large number of isomers. In addition, these systems tend to prefer non-symmetric structures with the Na and the electron pushed to one side of the cluster to minimize the perturbation of the hydrogen bond network. Given these facts, it is thus not so surprising that the clearest experimental result for a magic number cluster was found for Na-doped dimethyl ether clusters, namely for the hexamer Na(CH$_3$OCH$_3$)$_6$ which has near T$_h$ symmetry with an octahedral coordination of Na by the CH$_3$OCH$_3$ molecules. The lack of strong hydrogen bonding in these clusters strongly reduced the number of isomers, and in addition the hexamer was also found to be a particularly stable structure, i.e. it is also a magic number cluster with respect to stability. High level quantum chemical calculations for Na(CH$_3$OCH$_3$)$_n$ and Na(NH$_3$)$_n$ clusters by Gunina and Krylov[11] are in agreement with our previous experimental results[10, 24] and provide a detailed understanding of the underlying phenomena regarding the character of the electronic structure and the influence of structural fluctuations on the electronic properties.

The present study focuses on magic numbers in the photoelectron anisotropy of Li-doped dimethyl ether clusters (Li(CH$_3$OCH$_3$)$_n$). Many aspects of Li-doped molecular clusters have been investigated in detail (see refs.[9, 30-36] and references therein) but to the best of our knowledge no angle-resolved photoelectron spectra have been reported so far. Li is smaller and less polarizable than Na, which, for example, lets one expect that the almost perfect T$_h$ symmetry with octahedral coordination of the Na core in Na(CH$_3$OCH$_3$)$_6$ might be distorted in the Li-doped hexamer so that the magic number cluster might shift to another cluster size than the hexamer. The goal of the present work is to unravel how the substitution of the alkali metal in dimethyl ether clusters influences the energetics, structure, and magic numbers by a combination of experimental data and density functional theory (DFT) calculations.

## 2. Experiment

The experimental setup, the measurement procedures, and the data analysis are essentially identical to those used in our previous investigations of Na(CH$_3$OCH$_3$)$_n$ clusters.[10, 24] For convenience, we repeat here the main aspects as provided in the experimental part of West et al.[10] The experimental setup has been previously described in detail.[21-24, 37-39]

All measurements were performed in a velocity map imaging (VMI)[40, 41] photoelectron spectrometer, which can also function as a time-of-flight mass spectrometer. $(CH_3OCH_3)_n$ solvent clusters were generated by pulsed supersonic expansion of a He/$CH_3OCH_3$ gas mixture into vacuum. The solvent cluster size was varied from one molecule up to a maximum of approximately 175 molecules per cluster by varying the expansion conditions (backing pressure, gas composition, pressure, nozzle temperature) and oven temperature. The solvent clusters were doped with a single Li atom in a Li oven, which was heated to a temperature of 650 K. The resulting Li$(CH_3OCH_3)_n$ clusters were ionized with a 266 nm pulse from an Nd:YAG laser (photon energy of 4.66 eV), which exclusively ionized the unpaired (solvated) electron. The cluster size distributions were determined by mass spectrometry, which through the cluster mass provides information on the number of solvent molecules n per cluster.[21, 24, 37] For small clusters (n≤4) we use the actual number of molecules n to assign a cluster size, while the cluster size distributions for large clusters are characterized by the average cluster size <n> (and sometimes in addition by the maximum cluster size $n_{max}$). As exemplified by the relatively high intensity of the Na$(CH_3OCH_3)_6$ mass peak in figure 2f of West et al.[10] Relative intensities of clusters of different size in mass spectra can provide information on cluster stability (magic numbers related to cluster stability).

Information on cluster size dependent photoelectron angular distributions (PAD) and photoelectron kinetic energies are retrieved from the velocity map photoelectron images after reconstruction with MEVIR[42] (note that reconstruction with pBASEX[43] provides very similar results). Experimental electron binding energy (eBE) spectra are determined from the difference between the photon energy (4.66 eV) and the recorded photoelectron kinetic energy spectrum. The experimental ionization energies $IE_{max}$ for different cluster sizes are determined at the maxima of the photoelectron bands using Gaussian/Lorentzian fits. It is generally assumed that the $IE_{max}$ lie close to the values for the calculated vertical ionization energies $IE_{vert}$ (section 3). We characterize the PAD by the anisotropy parameter $\beta$,[44]

$$\frac{d\sigma}{d\Omega} = \frac{\sigma_{tot}}{4\pi}\left[1 + \frac{\beta}{2}\left(3\cos^2\theta - 1\right)\right], \qquad \text{Eq. (1)}$$

$\frac{d\sigma}{d\Omega}$ and $\sigma_{tot}$ are the differential and the total photoionization cross section, respectively, and $\theta$ is the angle between the photoelectron velocity vector and the polarization axis of the incident light. The indicated experimental cluster size dependent $\beta$-parameters are determined from an average over 11 pixels, including 5 pixels on each side of the peak maximum in the eBE spectra. As in our previous study,[10] we estimate the relative uncertainty of $IE_{max}$ and $\beta$ as a function of cluster size to be on the order of 5 % in both cases. The absolute uncertainties in $IE_{max}$ and $\beta$ are on the order of ± 0.1 eV and ± 0.1, respectively.

## 3. DFT calculations

The experimental results are compared with various quantities ($\beta$-parameters, vertical ionization energies $IE_{vert}$, enthalpies of vaporization $H_{vap}$, and dipole moments) obtained from calculations with the Gaussian program package[45] using the dispersion corrected ωB97XD density functional with a 6-31+G* basis set. The calculations are analogous to those for $Na(CH_3OCH_3)_n$ clusters,[10, 24] the most important aspects of which we repeat here for convenience. $H_{vap}$ is calculated for the neutral clusters as the total dissociation energy divided by the number of solvent monomer units. It is used here to compare cluster stabilities for different cluster sizes. The calculated total dipole moment of the different neutral clusters is used as a simple but very sensitive measure of the displacement of the charge distribution. $IE_{vert}$ are compared with the experimental $IE_{max}$. $IE_{vert}$ are obtained by subtracting the energy of the neutral cluster from the energy of the ionic cluster with the same geometry. The calculated $\beta$-parameters are determined as previously explained in detail[10] and in the supporting information of West et al.[24] Briefly, the highest occupied molecular orbital (HOMO) is expanded in terms of atomic natural orbitals (ANOs).[46] In order to account for the polarization of the HOMO upon solvation we use an expanded valence shell including 2p functions on Li for the ANO analysis (NBO program version 3.1). The normalized angular momentum ($\ell$) character $c_\ell^2$ of the HOMO is calculated as the sum over ANO contributions of the same $\ell$. The $\beta$-parameters are then obtained from,[17]

$$\beta = \sum_\ell c_\ell^2 \beta_\ell \qquad \text{Eq. (2)}$$

with $\beta_\ell$ determined from the Cooper–Zare formula,[44]

$$\beta_\ell = \frac{\ell(\ell-1)(1-R)^2 + (\ell+1)(\ell+2)R^2 - 6\ell(\ell+1)(1-R)R}{(2\ell+1)[\ell(1-R)^2 + (\ell+1)R^2]}. \qquad \text{Eq. (3)}$$

$R$ is the relative radial dipole matrix element of the ($\ell$+1) partial wave. We neglect the phase shift between outgoing partial waves. Furthermore, we provide here the results for $R = 0.5$ and for radial matrix elements that vanish at all centers except at the Li atom. We have previously shown for Na-doped clusters that the size-dependence of $\beta$ (not the actual values) is almost independent of the choice of the parameters (i.e. other limiting cases for $R$ and for radial matrix elements for all atomic centers).[24] In Li-doped and Na-doped clusters, the unpaired electron largely retains the character of the Li and Na valence electron, respectively. The above-mentioned robustness with respect to the model parameters derives from precisely this special property of the unpaired electron in the clusters, and enables us to derive meaningful results from the simple approach in Eq. (2) and (3). Note that trends in the size-dependence of calculated and experimental $\beta$-parameters can be compared, even though their actual values cannot.

## 4. Results for Li(CH$_3$OCH$_3$)$_n$ clusters

As an example Figure 1 shows a photoelectron image for small Li(CH$_3$OCH$_3$)$_n$ clusters with n≤4 together with the corresponding energy-dependent $\beta$-trace (full black line) and the photoelectron spectrum (dotted red line) as a function of the eBE. For these small clusters, the different rings in the image, the different bands in the energy-dependent $\beta$-trace, and the resolved bands in the photoelectron spectrum can be assigned to specific cluster sizes. The image and the $\beta$-trace show that the PAD remains clearly anisotropic (large values of the $\beta$-parameters at the band maxima), while the photoelectron spectrum reveals a very strong decrease in IE$_{max}$ by around 2eV with increasing cluster size from n=1 to n=4. For Li-doped clusters, truly size-resolved data could only be obtained up to n=4. As shown in Figure 2a, the photoelectron bands of larger clusters lie too close in energy for the current experiment to resolve specific cluster sizes. We thus assign an average cluster size <n> to these merged bands (see bands for <n>=20 and <n>=63 in Figure 2a). Figures 2b and 2c show representative mass spectra for the cases n=1-4 and <n>=63, respectively. The decrease in IE$_{max}$ with increasing cluster size is systematic but rather moderate beyond n=4 (see Figures 3g and 4b and Table S1 in the Supplementary Information). The evolution of the $\beta$-parameter with increasing cluster size is more complicated. Its value strongly decreases from $\beta$=1.4 to $\beta$=0.8 between n=1 and n=3, to peak again at n=4 with a value of $\beta$=1.3. Beyond n=4 we lose cluster size resolved information. However, the fact that $\beta$ stays fairly high up to <n>=20 (see Figures 3e and 4c and Table S1 in the Supplementary Information) implies that *several* clusters with sizes larger than n=4 must also have fairly high *β*-parameters. At first sight, this result appears rather surprising.

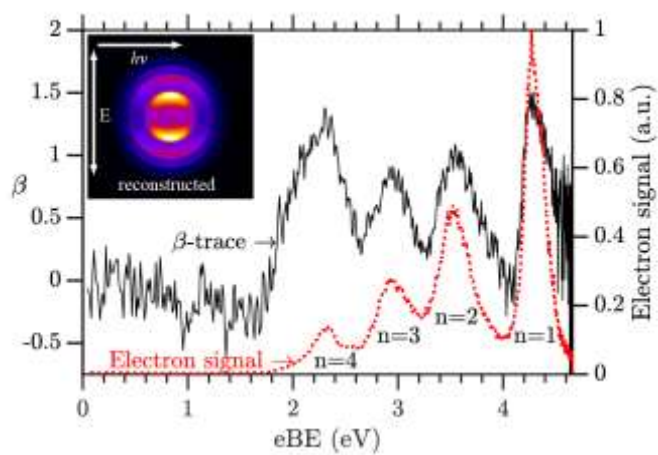

**Figure 1:** Inset: reconstructed photoelectron images of Li(CH$_3$OCH$_3$)$_n$ clusters with n=1-4 solvent molecules. Full black line: *β*-trace as a function of the electron binding energy (eBE). Dotted red line: Photoelectron spectrum as a function of eBE.

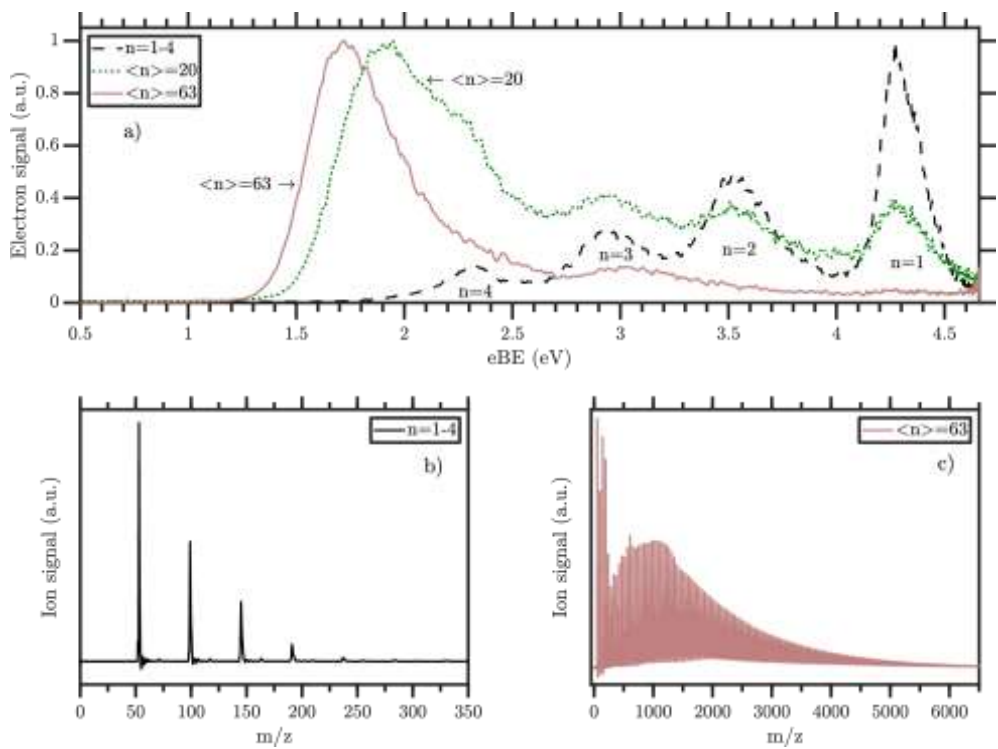

**Figure 2:** a) Photoelectron spectra of Li(CH$_3$OCH$_3$)$_n$ clusters as a function of eBE. Dashed black line: n=1-4; Dotted green line: <n>=20; Full red line: <n>=63. b) Mass spectrum for n=1-4. c) Mass spectrum for <n>=63.

To better understand this behavior, we have performed DFT calculations to obtain structures and *β*-parameters for clusters with up to 20 molecules (see Figure 3 and Table S2 in the Supplementary Information). The calculated *β*-parameter of 2 in Figure 3a and the isosurface of the HOMO in Figure 5a clearly reveal the tetramer Li(CH$_3$OCH$_3$)$_4$ as a magic cluster with respect to the photoelectron anisotropy consistent with the experiment (Figure 3e). This cluster has a highly symmetric structure with a tetrahedral coordination of the Li core by the ligands and a symmetric highest occupied molecular orbital (HOMO) that is delocalized over an almost spherical shell around the cluster (Figure 5a). As a result, this HOMO has a very high s-character (100% in the calculation), which explains the high *β* value. The high symmetry is also reflected in the very low dipole moment (Figure 3d). However, H$_{vap}$ in Figure 3b shows that Li(CH$_3$OCH$_3$)$_4$ is not particularly stable compared with the neighboring cluster sizes; i. e. it is not a magic cluster with respect to the stability. This is consistent with the experimental mass spectra, in which the mass signal does not peak at n=4 (see example in Figures 2c and 3f). In the case of sodium clusters we had observed a rather different behavior, with the Na-hexamer Na(CH$_3$OCH$_3$)$_6$ as a magic cluster with respect to both stability and anisotropy (see figures 2b and 2f in West et al.[10]). This difference between Li and Na can be rationalized by the balance between the electronic stabilization afforded by the metal-oxygen "bonds" on one hand and the steric destabilization due to crowding of ligands on the other hand. In small Na clusters the former dominates, while the latter gains importance in Li clusters as a result of the much smaller atomic radius and hence shorter metal-oxygen bond (see further below and section 5). Li(CH$_3$OCH$_3$)$_4$ shows a highly symmetric coordination with an almost perfect LiO$_4$ tetrahedron. The T$_d$ symmetry is necessarily broken by the C$_2$ symmetry of the ligands resulting in a number of symmetry equivalent minima. At the ωB97XD/6-31+G* level the rotation angles around the C$_3$-axes of the LiO$_4$ tetrahedron are virtually all equal, resulting in a highly symmetric HOMO. As a result the dipole moment almost vanishes and $\beta$ reaches its maximum value of 2. At higher levels of electronic theory the symmetry of the equilibrium structure might be further broken, resulting in the localization of the unpaired electron at on side of the (approximate) tetrahedron, a correspondingly large dipole moment and a reduction of the calculated $\beta$. Nevertheless one would still expect to observe tetrahedral symmetry of the HOMO in the experiment. This is a consequence of the inverse Born-Oppenheimer (or sudden) character of the unpaired electron's wave function: very small nuclear displacements lead to a large change in the electronic wave function. Effectively, the unpaired electron will "see" a vibrationally averaged structure of the cluster. As long as the barrier between symmetry-equivalent minima remains small (as expected for small rotation angles around the $C_3$-axes of the $LiO_4$ tetrahedron), the vibrationally averaged structure will remain tetrahedral.

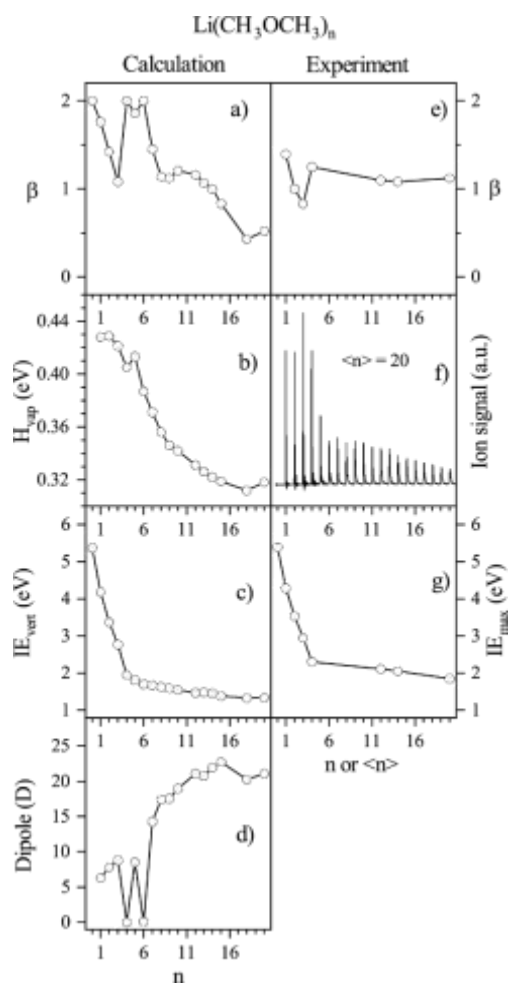

**Figure 3:** Properties of Li(CH$_3$OCH$_3$)$_n$ clusters as a function of the number of solvent molecules n: a) calculated $\beta$-parameters, b) calculated enthalpies of vaporization H$_{vap}$, c) calculated vertical ionization energies IE$_{vert}$, d) calculated dipole moments, e) experimental $\beta$-parameters, f) representative mass spectrum for <n>=20, and g) experimental ionization energies determined at the maximum of the photoelectron bands IE$_{max}$. For the calculations, the open circles connected by lines are the values for the energetically lowest isomers. For other isomers see Table S2 in the Supplementary Information.

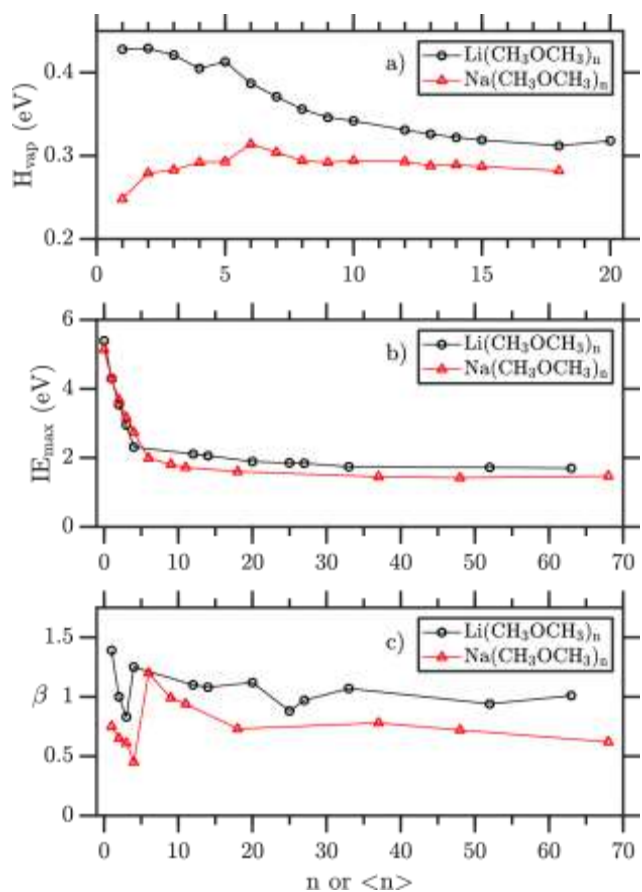

**Figure 4:** Comparison of Li(CH$_3$OCH$_3$)$_n$ (open circles) and Na(CH$_3$OCH$_3$)$_n$ (open triangles) cluster data a) Calculated enthalpies of vaporization H$_{vap}$ b) Experimental ionization energies IE$_{max}$ c) Experimental $\beta$-parameters.

The next two larger cluster sizes, Li(CH$_3$OCH$_3$)$_5$ and Li(CH$_3$OCH$_3$)$_6$, have similarly high $\beta$-parameters as the tetramer (Figure 3a). This is rather surprising and different from the Na-case in West et al.[10] However, it supports our hypothesis that several clusters with sizes larger than n=4 and high $\beta$-parameters are the reason for the experimentally observed trend of high $\beta$-values even up to <n>=20 (Figure 3e) because the contribution of the highly symmetric smaller cluster is high in the corresponding size distributions. The most stable pentamer has a trigonal-bipyramidal structure (Figure 5b) and the most stable hexamer has an octahedral structure (Figure 5c). To make space for the additional ligands in the expanded solvation shell the Li-O distance increases. This weakening of the Li-O "bond" reduces $H_{vap}$. In the case of the pentamer, an axial distortion of the trigonal-bipyramidal structure counteracts this effect by reducing the unfavorable steric interaction between the ligands, so that overall $H_{vap}$ even slightly increases compared with the tetramer (but still lies significantly below the trimer). The crowding of ligands leads to an additional symmetry breaking in terms of the rotational angles of the three equatorial ligands around the Li-O "bonds". The ensuing localization of the unpaired electron on one side of the pyramid is reflected in the sizeable dipole moment. The dominant s-character of the HOMO (95%) is much less affected by this localization with a correspondingly small reduction of the $\beta$ value. In the hexamer cluster the further lengthening of the Li-O distance is the dominant effect, such that $H_{vap}$ decreases significantly. The octahedral structure remains intact with a highly symmetric HOMO (100% s-character) and correspondingly vanishing dipole moment and maximal $\beta$ value. All these clusters correspond to the completion of a first coordination shell, so that they can be considered as "isomeric motifs of the first coordination shell". This explains their similar values for $\beta$ and $IE_{vert}$ and their relatively high stability (high values of $H_{vap}$). Table S2 in the Supplementary Information also lists data for a few isomers for n=5 and 6. Isomer (a) of the pentamer has a tetrahedral first ligand shell with one additional molecule added in the 2$^{nd}$ shell. Its $\beta$ value of 1.49 is significantly lower than that for the most stable isomer because of the lower symmetry. Similarly, isomers of the hexamer with a symmetric first ligand shell and additional molecules placed in the 2$^{nd}$ shell also have low $\beta$ values (isomer (b) with a first tetrahedral shell $\beta$=1.34 and isomer (a) with a first bipyramidal ligand shell $\beta$=1.44, see Table S2).

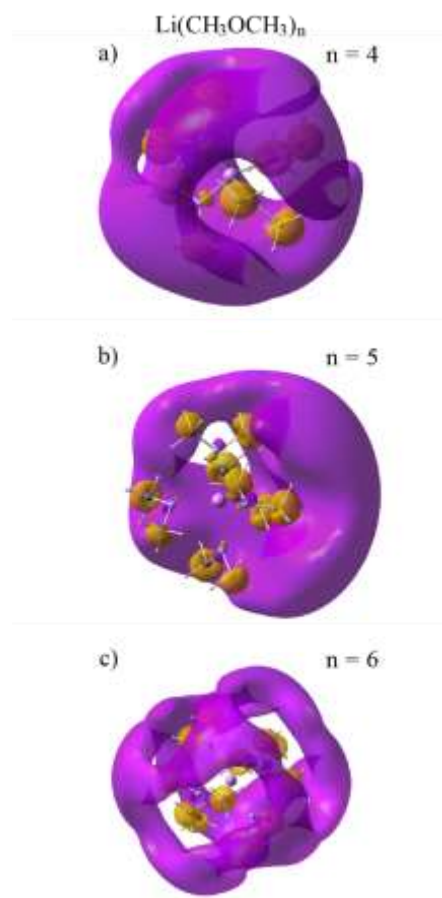

**Figure 5:** Isosurfaces of the HOMO of the most stable isomers of a) the Li(CH$_3$OCH$_3$)$_4$ cluster, b) the Li(CH$_3$OCH$_3$)$_5$ cluster, and c) the Li(CH$_3$OCH$_3$)$_6$ cluster. The calculated s-character of these HOMO are 100%, 95%, and 100%, respectively.

The PADs of even larger clusters are still anisotropic but with $\beta$ values clearly below those of n=4,5 and 6 (Figure 3a). These clusters are less symmetric than the smaller ones with correspondingly lower $\beta$ values and larger dipoles. The less symmetric structures – typically with the Li and its electron pushed to one side of the cluster (see Figure S1 in the Supporting Information for n=20) allow the perturbation of the solvent molecules to be minimized, while keeping the unpaired electron close to the Li core and maximizing the strong favorable Li-O interactions in the 1$^{st}$ solvation shell. Table S2 in the Supplementary Information lists the properties of some higher lying isomers. Among them are also highly symmetric isomers with high $\beta$ values, such as isomer (b) for n=10. For Li-clusters, H$_{vap}$ decreases almost continuously with increasing cluster size (Figure 3b). As expected, for very large clusters it converges to the calculated bulk value of pure (without Li) dimethyl ether of about 0.23eV (experimental bulk value around 0.29eV).[47] For small clusters with one ligand shell (up to n~6), $H_{vap}$ is comparatively high because the strong Li-O bond dominates – partially counterbalanced by the steric interaction in the increasingly crowded ligand shell. With more ligands (n≳7), the contribution of the Li-O bonds to $H_{vap}$ is increasingly 'diluted' by the much weaker ligand-ligand interaction in the outer shells and gradually converges to the bulk value. For $IE_{vert}$, pronounced changes are only observed until the completion of the first solvation shell at n=4. The extension of the 1$^{st}$ shell in n=5 and 6 retains a balance between the increase in the number of strong Li-O interactions and their weakening as a consequence of ligand crowding (bond lengthening). The further slow decrease of $IE_{vert}$ beyond n=6 can be attributed to increasing polarization effects (as the cluster's polarizability increases with its size).

## 5. Comparison of Li(CH$_3$OCH$_3$)$_n$ and Na(CH$_3$OCH$_3$)$_n$ clusters

Figure 4 provides a comparison of Li(CH$_3$OCH$_3$)$_n$ and Na(CH$_3$OCH$_3$)$_n$ cluster data. The behavior of $H_{vap}$ for the Na-clusters differs pronouncedly from that of the Li-clusters (Figure 4a). Small Na-clusters have a lower $H_{vap}$ that increases with cluster size, while small Li clusters have a higher $H_{vap}$ that decreases with cluster size. The maxima for $H_{vap}$ are reached at the hexamer of Na and at the monomer for Li. The generally lower $H_{vap}$ for small Na-clusters can likely be attributed to the weaker Na-O bond compared with the Li-O bond. As mentioned in section 4, the decrease of $H_{vap}$ is consistent with a weakening (i.e. lengthening) of the Li-O bond because of the increased crowding of ligands in the 1$^{st}$ solvation shell. Given the much larger atomic radius of sodium ligand crowding plays a less important (if any) role in small Na(CH$_3$OCH$_3$)$_n$ clusters. This would lead to the expectation of a roughly constant $H_{vap}$ until the 1st solvation shell is complete at n=6. The increase of $H_{vap}$ observed instead points toward significant cooperative effects, possibly resulting in part from (weak) hydrogen bonding interactions between the ligands. For larger clusters (n>6), $H_{vap}$ decreases again, but more slowly than for Li-clusters. This is in part a trivial consequence of the smaller difference between the Na-O bond strength and the ligand-ligand interaction (the "dilution" per ligand added is less in Na- than in Li-clusters). Another contributing factor is the larger polarizability of the 3s unpaired electron of Na as compared with the 2s electron of Li. The former more easily deforms to adapt to its position on the cluster surface.

In contrast to the trends in $H_{vap}$, the trends in $IE_{max}$ are qualitatively identical for Na- and Li-clusters (Figure 4b). Strong decreases are only observed before the closure of the first solvation shells (at n=4 for Li and n=6 for Na), while for larger clusters the values of $IE_{max}$ drop only very slowly as a result of the increasing overall polarizability of the cluster. For larger Li clusters, the absolute values of $IE_{max}$ observed experimentally lie systematically above those of the Na clusters by about 0.3 eV. This difference approximately equals the difference between the ionization energies for atomic Na and Li (5.14 eV[48] and 5.39 eV,[49] respectively). The unpaired (surface-solvated) electron in the cluster apparently still feels the core it belonged to. This is consistent with the results of our DFT calculations. As mentioned above, the most stable larger clusters tend to have the metal core and the electron located at one side of the cluster close to the surface. The electron is thus still close to the respective metal core, which might explain the conservation of the shift between cluster and atomic metal.

Finally, Figure 4c compares the $\beta$-parameters for the two cases. The occurrence of magic clusters related to the anisotropy at n=6 for Na clusters and at n=4,5, and 6 for Li clusters was already discussed in West et al.[10] and in section 4. Here, we additionally point out the general downshift of $\beta$ of Na-clusters compared with Li-clusters observed experimentally for essentially all cluster sizes. This phenomenon is reproduced at least qualitatively by the DFT calculations. A lowering of $β$ results from the polarization of the HOMO upon solvation, which gives rise to higher angular momentum components $l$ (essentially $l = 1$) of the HOMO. A more polarizable atomic orbital is more easily distorted (polarized), i.e. more easily acquires higher $l$ components upon solvation of the atom. The lower $β$ values of the Na-clusters can thus be explained by the higher polarizability of the 3s electron compared with the 2s electron of Li.

## 6. Summary

This paper compares properties of neutral $Li(CH_3OCH_3)_n$ and $Na(CH_3OCH_3)_n$ clusters with a focus on magic numbers related to the photoelectron anisotropies of the highest occupied molecular orbital; i. e. the solvated electron which can delocalized over extended cluster regions. In Li-doped clusters, magic numbers are observed at n=4,5, and 6 as a result of the completion of the first solvation shell. Such "isomeric motifs of the first coordination shell" were not observed for Na-doped clusters, which showed a distinct magic cluster at n=6.[10] The difference between the two alkali metals seems to arise from a balance between the electronic stabilization by the metal-oxygen bonds and the steric destabilization due to crowding of ligands. The general lowering of the $β$ parameters of around 0.25 for Na-clusters compared with Li-clusters for larger clusters with up to <n>≈70 can be explained by the higher polarizability of the 3s compared with the 2s electron. Similarly, a general lowering of the ionization energy by approximately 0.3 eV of Na-clusters compared with Li-clusters is observed in the same cluster size range. It roughly matches the difference of the ionization energies of the two bare metals, which seems to be conserved in the molecular clusters. Both alkali metal clusters show a very pronounced decrease of the ionization energy by about 2 eV for small clusters before the closure of the first solvation shell. DFT calculations reveal a distinct difference between the behavior of the enthalpies of vaporization for the two metal clusters as a function of the cluster size, which can be rationalized by differences in the metal-oxygen bonds and the ligand-ligand interactions, in the polarizabilities of the 3s and 2s electrons, in the crowding of ligands, and by cooperative effects. Such cluster studies might also contribute to a better understanding of the properties of the solvated electron in the condensed phase.[9, 50-61]

**ACKNOWLEDGMENT**

Financial support was provided by the Swiss National Science Foundation under project no. 200020_172472 and by the ETH Zürich. This project has received funding from the European Union's Horizon 2020 research and innovation program from the European Research Council under the Grant Agreement No 786636. We are very grateful to David Stapfer and Markus Steger from our workshop for their help in the setup of the Li oven, and to Dr. David Luckhaus for his help with the calculations.

**REFERENCES**

(1) Vafayi, K.; Esfarjani, K. Abundance of Nanoclusters in a Molecular Beam: The Magic Numbers for Lennard-Jones Potential. *J. Cluster Sci.* **2015**, *26*, 473-490.
(2) Coolbaugh, M. T.; Garvey, J. F. Magic Numbers in Molecular Clusters - a Probe for Chemical-Reactivity. *Chem. Soc. Rev.* **1992**, *21*, 163-169.
(3) Alonso, J. A. *Structure and Properties of Atomic Nanoclusters*; Imperial College Press: London, 2005.
(4) Chang, H. C.; Wu, C. C.; Kuo, J. L. Recent Advances in Understanding the Structures of Medium-Sized Protonated Water Clusters. *Int. Rev. Phys. Chem.* **2005**, *24*, 553-578.
(5) Young, R. M.; Neumark, D. M. Dynamics of Solvated Electrons in Clusters. *Chem. Rev.* **2012**, *112*, 5553-5577.
(6) Hammer, N. I.; Shin, J. W.; Headrick, J. M.; Diken, E. G.; Roscioli, J. R.; Weddle, G. H.; Johnson, M. A. How Do Small Water Clusters Bind an Excess Electron? *Science* **2004**, *306*, 675-679.
(7) Turi, L.; Rossky, P. J. Theoretical Studies of Spectroscopy and Dynamics of Hydrated Electrons. *Chem. Rev.* **2012**, *112*, 5641-5674.
(8) Zeuch, T.; Buck, U. Sodium Doped Hydrogen Bonded Clusters: Solvated Electrons and Size Selection. *Chem. Phys. Lett.* **2013**, *579*, 1-10.
(9) Zurek, E.; Edwards, P. P.; Hoffmann, R. A Molecular Perspective on Lithium-Ammonia Solutions. *Angew. Chem. Int. Ed.* **2009**, *48*, 8198-8232.
(10) West, A. H. C.; Yoder, B. L.; Luckhaus, D.; Signorell, R. Solvated Electrons in Clusters: Magic Numbers for the Photoelectron Anisotropy. *J. Phys. Chem. A* **2015**, *119*, 12376-12382.
(11) Gunina, A. O.; Krylov, A. I. Probing Electronic Wave Functions of Sodium-Doped Clusters: Dyson Orbitals, Anisotropy Parameters, and Ionization Cross-Sections. *J. Phys. Chem. A* **2016**, *120*, 9841-9856.
(12) Borgis, D.; Rossky, P. J.; Turi, L. Electronic Excited State Lifetimes of Anionic Water Clusters: Dependence on Charge Solvation Motif. *J. Phys. Chem. Lett.* **2017**, *8*, 2304-2309.
(13) Bartels, C.; Hock, C.; Huwer, J.; Kuhnen, R.; Schwöbel, J.; von Issendorff, B. Probing the Angular Momentum Character of the Valence Orbitals of Free Sodium Nanoclusters. *Science* **2009**, *323*, 1323-1327.
(14) Bartels, C. Angular Distributions of Photoelectrons from Cold, Size-Selected Sodium Cluster Anions. Ph.D. Dissertation, Albert-Ludwigs-Universität, Freiburg im Breisgau, Germany, 2008.
(15) Rolles, D.; Zhang, H.; Pešić, Z. D.; Bilodeau, R. C.; Wills, A.; Kukk, E.; Rude, B. S.; Ackerman, G. D.; Bozek, J. D.; Díez Muiño, R.; et al. Size Effects in Angle-Resolved Photoelectron Spectroscopy of Free Rare-Gas Clusters. *Phys. Rev. A* **2007**, *75*, 031201.
(16) Sanov, A. Laboratory-Frame Photoelectron Angular Distributions in Anion Photodetachment: Insight into Electronic Structure and Intermolecular Interactions. *Annu. Rev. Phys. Chem.* **2014**, *65*, 341-363.
(17) Melko, J. J.; Castleman, A. W., Jr. Photoelectron Imaging of Small Aluminum Clusters: Quantifying s-p Hybridization. *Phys. Chem. Chem. Phys.* **2013**, *15*, 3173-3178.
(18) Khuseynov, D.; Blackstone, C. C.; Culberson, L. M.; Sanov, A. Photoelectron Angular Distributions for States of Any Mixed Character: An Experiment-Friendly Model for Atomic, Molecular, and Cluster Anions. *J. Chem. Phys.* **2014**, *141*, 124312.
(19) Kammrath, A.; Verlet, J. R. R.; Griffin, G. B.; Neumark, D. M. Photoelectron Imaging of Large Anionic Methanol Clusters: $(MeOH)_n^-$ (n ~ 70-460). *J. Chem. Phys.* **2006**, *125*, 171102.

(20) Young, R. M.; Yandell, M. A.; Niemeyer, M.; Neumark, D. M. Photoelectron Imaging of Tetrahydrofuran Cluster Anions (THF)$_n^-$ (1 ≤ n ≤ 100). *J. Chem. Phys.* **2010**, *133*, 154312.

(21) Signorell, R.; Yoder, B. L.; West, A. H. C.; Ferreiro, J. J.; Saak, C.-M. Angle-Resolved Valence Shell Photoelectron Spectroscopy of Neutral Nanosized Molecular Aggregates. *Chem. Sci.* **2014**, *5*, 1283-1295.

(22) West, A. H. C.; Yoder, B. L.; Signorell, R. Size-Dependent Velocity Map Photoelectron Imaging of Nanosized Ammonia Aerosol Particles. *J. Phys. Chem. A* **2013**, *117*, 13326-13335.

(23) Yoder, B. L.; West, A. H. C.; Schläppi, B.; Chasovskikh, E.; Signorell, R. A Velocity Map Imaging Photoelectron Spectrometer for the Study of Ultrafine Aerosols with a Table-Top VUV Laser and Na-Doping for Particle Sizing Applied to Dimethyl Ether Condensation. *J. Chem. Phys.* **2013**, *138*, 044202.

(24) West, A. H. C.; Yoder, B. L.; Luckhaus, D.; Saak, C.-M.; Doppelbauer, M.; Signorell, R. Angle-Resolved Photoemission of Solvated Electrons in Sodium-Doped Clusters. *J. Phys. Chem. Lett.* **2015**, *6*, 1487-1492.

(25) Oana, C. M.; Krylov, A. I. Dyson Orbitals for Ionization from the Ground and Electronically Excited States within Equation-of-Motion Coupled-Cluster Formalism: Theory, Implementation, and Examples. *J. Chem. Phys.* **2007**, *127*, 234106.

(26) Oana, C. M.; Krylov, A. I. Cross Sections and Photoelectron Angular Distributions in Photodetachment from Negative Ions Using Equation-of-Motion Coupled-Cluster Dyson Orbitals. *J. Chem. Phys.* **2009**, *131*, 124114.

(27) Yamamoto, Y.-I.; Suzuki, Y.-I.; Tomasello, G.; Horio, T.; Karashima, S.; Mitrić, R.; Suzuki, T. Time- and Angle-Resolved Photoemission Spectroscopy of Hydrated Electrons near a Liquid Water Surface. *Phys. Rev. Lett.* **2014**, *112*, 187603.

(28) Peppernick, S. J.; Gunaratne, K. D. D.; Castleman, A. W., Jr. Superatom Spectroscopy and the Electronic State Correlation between Elements and Isoelectronic Molecular Counterparts. *Proc. Natl. Acad. Sci. U.S.A.* **2010**, *107*, 975-980.

(29) Humeniuk, A.; Wohlgemuth, M.; Suzuki, T.; Mitrić, R. Time-Resolved Photoelectron Imaging Spectra from Non-Adiabatic Molecular Dynamics Simulations. *J. Chem. Phys.* **2013**, *139*, 134104.

(30) Hopkins, W. S.; Woodham, A. P.; Tonge, N. M.; Ellis, A. M.; Mackenzie, S. R. Photodissociation Dynamics of Li(NH$_3$)$_4$.: A Velocity Map Imaging Study. *J. Phys. Chem. Lett.* **2011**, *2*, 257-261.

(31) Varriale, L.; Tonge, N. M.; Bhalla, N.; Ellis, A. M. Communications: The Electronic Spectrum of Li(NH$_3$)$_4$. *J. Chem. Phys.* **2010**, *132*, 161101.

(32) Salter, T. E.; Ellis, A. M. Microsolvation of Lithium in Ammonia: Dissociation Energies and Spectroscopic Parameters of Small Li(NH$_3$)$_n$ Clusters (n=1 and 2) and Their Cations. *Chem. Phys.* **2007**, *332*, 132-138.

(33) Salter, T. E.; Mikhailov, V. A.; Evans, C. J.; Ellis, A. M. Infrared Spectroscopy of Li(NH$_3$)$_n$ Clusters for n=4-7. *J. Chem. Phys.* **2006**, *125*, 034302.

(34) More, M. B.; Glendening, E. D.; Ray, D.; Feller, D.; Armentrout, P. B. Cation-Ether Complexes in the Gas Phase: Bond Dissociation Energies and Equilibrium Structures of Li$^+$[O(CH$_3$)$_2$]$_x$, x=1-4. *J. Phys. Chem.* **1996**, *100*, 1605-1614.

(35) Sohnlein, B. R.; Li, S. G.; Fuller, J. F.; Yang, D. S. Pulsed-Field Ionization Electron Spectroscopy and Binding Energies of Alkali Metal Dimethyl Ether and -Dimethoxyethane Complexes. *J. Chem. Phys.* **2005**, *123*, 014318.

(36) Takasu, R.; Hashimoto, K.; Fuke, K. Study on Microscopic Solvation Process of Li Atom in Ammonia Clusters: Photoionization and Photoelectron Spectroscopies of M(NH$_3$)$_n$ (M=Li, Li$^-$). *Chem. Phys. Lett.* **1996**, *258*, 94-100.

(37) Yoder, B. L.; Litman, J. H.; Forysinski, P. W.; Corbett, J. L.; Signorell, R. Sizer for Neutral Weakly Bound Ultrafine Aerosol Particles Based on Sodium Doping and Mass Spectrometric Detection. *J. Phys. Chem. Lett.* **2011**, *2*, 2623-2628.

(38) Litman, J. H.; Yoder, B. L.; Schläppi, B.; Signorell, R. Sodium-Doping as a Reference to Study the Influence of Intracluster Chemistry on the Fragmentation of Weakly-Bound Clusters Upon Vacuum Ultraviolet Photoionization. *Phys. Chem. Chem. Phys.* **2013**, *15*, 940-949.

(39) Forysinski, P. W.; Zielke, P.; Luckhaus, D.; Corbett, J.; Signorell, R. Photoionization of Small Sodium-Doped Acetic Acid Clusters. *J. Chem. Phys.* **2011**, *134*, 094314.

(40) Chandler, D. W.; Houston, P. L. Two-Dimensional Imaging of State-Selected Photodissociation Products Detected by Multiphoton Ionization. *J. Chem. Phys.* **1987**, *87*, 1445-1447.

(41) Eppink, A.; Parker, D. H. Velocity Map Imaging of Ions and Electrons Using Electrostatic Lenses: Application in Photoelectron and Photofragment Ion Imaging of Molecular Oxygen. *Rev. Sci. Instrum.* **1997**, *68*, 3477-3484.

(42) Dick, B. Inverting Ion Images without Abel Inversion: Maximum Entropy Reconstruction of Velocity Maps. *Phys. Chem. Chem. Phys.* **2014**, *16*, 570-580.

(43) Garcia, G. A.; Nahon, L.; Powis, I. Two-Dimensional Charged Particle Image Inversion Using a Polar Basis Function Expansion. *Rev. Sci. Instrum.* **2004**, *75*, 4989-4996.

(44) Cooper, J.; Zare, R. N. Angular Distribution of Photoelectrons. *J. Chem. Phys.* **1968**, *48*, 942-943.

(45) Frisch, M. J.; Trucks, G. W.; Schlegel, H. B.; Scuseria, G. E.; Robb, M. A.; Cheeseman, J. R.; Scalmani, G.; Barone, V.; Mennucci, B.; Petersson, G. A.; et al. Gaussian 09. Wallingford, CT, USA: Gaussian, Inc.; 2009.

(46) Reed, A. E.; Weinstock, R. B.; Weinhold, F. Natural-Population Analysis. *J. Chem. Phys.* **1985**, *83*, 735-746.

(47) Acree, W. E.; Chickos, J. S. Phase Transition Enthalpy Measurements of Organic and Organometallic Compounds. In: Linstrom, P. J., Mallard, W. G., editors. Nist Chemistry Webbook, Nist Standard Reference Database Number 69. Gaithersburg MD: National Institute of Standards and Technology; **retreived December 18, 2018**.

(48) Peterson, K. I.; Dao, P. D.; Farley, R. W.; Castleman, A. W. Photoionization of Sodium Clusters. *J. Chem. Phys.* **1984**, *80*, 1780-1785.

(49) Bushaw, B. A.; Nörtershauser, W.; Drake, G. W. F.; Kluge, H. J. Ionization Energy of $^{6,7}$Li Determined by Triple-Resonance Laser Spectroscopy. *Phys. Rev. A* **2007**, *75*, 052503.

(50) Holton, D.; Edwards, P. Metals in Non-Aqueous Solvents. *Chem. Britain* **1985**, *21*, 1007-1013.

(51) Thomas, J. M.; Edwards, P. P.; Kuznetsov, V. L. Sir Humphry Davy: Boundless Chemist, Physicist, Poet and Man of Action. *ChemPhysChem* **2008**, *9*, 59-66.

(52) Seel, A. G.; Swan, H.; Bowron, D. T.; Wasse, J. C.; Weller, T.; Edwards, P. P.; Howard, C. A.; Skipper, N. T. Electron Solvation and the Unique Liquid Structure of a Mixed-Amine Expanded Metal: The Saturated Li-NH3-MeNH$_2$ System. *Angew. Chem. Int. Ed.* **2017**, *56*, 1561-1565.

(53) Hartweg, S.; West, A. H. C.; Yoder, B. L.; Signorell, R. Metal Transition in Sodium-Ammonia Nanodroplets. *Angew. Chem. Int. Ed.* **2016**, *55*, 12347-12350.

(54) Coons, M. P.; You, Z. Q.; Herbert, J. M. The Hydrated Electron at the Surface of Neat Liquid Water Appears to Be Indistinguishable from the Bulk Species. *J. Am. Chem. Soc.* **2016**, *138*, 10879-10886.

(55) Karashima, S.; Yamamoto, Y.; Suzuki, T. Resolving Nonadiabatic Dynamics of Hydrated Electrons Using Ultrafast Photoemission Anisotropy. *Phys. Rev. Lett.* **2016**, *116*, 137601.
(56) Luckhaus, D.; Yamamoto, Y. I.; Suzuki, T.; Signorell, R. Genuine Binding Energy of the Hydrated Electron. *Sci. Adv.* **2017**, *3*, e1603224.
(57) Elkins, M. H.; Williams, H. L.; Shreve, A. T.; Neumark, D. M. Relaxation Mechanism of the Hydrated Electron. *Science* **2013**, *342*, 1496-1499.
(58) Stähler, J.; Deinert, J. C.; Wegkamp, D.; Hagen, S.; Wolf, M. Real-Time Measurement of the Vertical Binding Energy During the Birth of a Solvated Electron. *J. Am. Chem. Soc.* **2015**, *137*, 3520-3524.
(59) Riley, J. W.; Wang, B. X.; Woodhouse, J. L.; Assmann, M.; Worth, G. A.; Fielding, H. H. Unravelling the Role of an Aqueous Environment on the Electronic Structure and Ionization of Phenol Using Photoelectron Spectroscopy. *J. Phys. Chem. Lett.* **2018**, *9*, 678-682.
(60) Savolainen, J.; Uhlig, F.; Ahmed, S.; Hamm, P.; Jungwirth, P. Direct Observation of the Collapse of the Delocalized Excess Electron in Water. *Nature Chem.* **2014**, *6*, 697-701.
(61) Kumar, G.; Roy, A.; McMullen, R. S.; Kutagulla, S.; Bradforth, S. E. The Influence of Aqueous Solvent on the Electronic Structure and Non-Adiabatic Dynamics of Indole Explored by Liquid-Jet Photoelectron Spectroscopy. *Faraday Discuss.* **2018**, *212*, 359-381.

# Supplementary Information:

# Magic Numbers for the Photoelectron Anisotropy in

# Li-Doped Dimethyl Ether Clusters

*Jonathan V. Barnes, Bruce L. Yoder, and Ruth Signorell\**

ETH Zürich, Laboratory of Physical Chemistry, Vladimir-Prelog-Weg 2, CH-8093, Zürich, Switzerland

\* To whom correspondence should be addressed. E-mail: rsignorell@ethz.ch

**Table S1**: Experimental Results of the reconstructed images using MEVIR. n, <n> and $n_{max}$ are the number of solvent molecules per cluster, the average number of solvent molecules per cluster and the maximum number of solvent molecules per cluster, respectively. $IE_{max}$ is the determined peak position for the individual cluster sizes in the photoelectron spectra and $\beta$ the anisotropy parameter.

| n | <n> | $n_{max}$ | $IE_{max}$ (eV) | $\beta$ |
|---|---|---|---|---|
| 0 |   |   | 5.39 | - |
| 1 |   |   | 4.30 | 1.39 |
| 2 |   |   | 3.54 | 1.00 |
| 3 |   |   | 2.95 | 0.83 |
| 4 |   |   | 2.31 | 1.25 |
|   | 12 | 18 | 2.11 | 1.10 |
|   | 14 | 21 | 2.06 | 1.08 |
|   | 20 | 52 | 1.89 | 1.12 |
|   | 25 | 63 | 1.85 | 0.88 |
|   | 27 | 70 | 1.84 | 0.97 |
|   | 33 | 83 | 1.74 | 1.07 |
|   | 52 | 142 | 1.72 | 0.94 |
|   | 63 | 175 | 1.70 | 1.01 |

**Table S2**: Results of the DFT calculations (wB97XD/6-31+G*) using Gaussian 09. Enthalpies of vaporization H$_{vap}$, dipole moments, vertical ionization energies IE$_{vert}$, and β-parameters. The table has been divided into two sections in order to display it more effectively.

| n | isomer | H$_{vap}$ (eV) | dipole (D) | IE$_{vert}$ (eV) | β |
|---|---|---|---|---|---|
| 0 | | - | - | 5.37 | 2.00 |
| 1 | | 0.428 | 6.3 | 4.19 | 1.76 |
| 2 | | 0.429 | 7.8 | 3.38 | 1.42 |
| 3 | | 0.421 | 8.8 | 2.77 | 1.08 |
| 4 | | 0.405 | 0.0 | 1.95 | 2.00 |
| 5 | | 0.413 | 8.5 | 1.81 | 1.86 |
|   | a | 0.384 | 13.1 | 1.82 | 1.49 |
| 6 | | 0.387 | 0.1 | 1.70 | 2.00 |
|   | a | 0.380 | 13.8 | 1.76 | 1.44 |
|   | b | 0.365 | 15.5 | 1.72 | 1.34 |
| 7 | | 0.371 | 14.3 | 1.68 | 1.45 |
|   | a | 0.363 | 16.0 | 1.70 | 1.21 |
| 8 | | 0.356 | 17.4 | 1.64 | 1.14 |
|   | a | 0.353 | 15.9 | 1.62 | 1.29 |
| 9 | | 0.346 | 17.5 | 1.60 | 1.12 |
|   | a | 0.344 | 16.3 | 1.60 | 1.19 |

n=2: bent.
n=3: flat trigonal pyramid.
n=4: tetrahedron.
n=5: trigonal bipyramid. (a) n=4 (tetrahedron) with one ligand added in the 2$^{nd}$ shell.
n=6: octahedron. (a): n=5 (trigonal bipyramid) with one ligand added in the 2$^{nd}$ shell. (b) n=4 (tetrahedron) with two ligands added in the 2$^{nd}$ shell.
n=7: n=6 (octahedron) with one ligands added in the 2$^{nd}$ shell. (a) n=5 (trigonal bipyramid) with two ligands added in the 2$^{nd}$ shell.
n=8: n=5 (trigonal bipyramid) with three ligands added in the 2$^{nd}$ shell. (a) n=6 (octahedron) with two ligands added in the 2$^{nd}$ shell.
n=9: n=5 (trigonal bipyramid) with four ligands added in the 2$^{nd}$ shell. (a) n=6 (octahedron) with three ligands added in the 2$^{nd}$ shell.

**Table S2 (continued)**: Results of the DFT calculations (wB97XD/6-31+G*) using Gaussian 09. Enthalpies of vaporization H$_{vap}$, dipole moments, vertical ionization energies IE$_{vert}$, and β-parameters. The table has been divided into two sections in order to display it more effectively.

| n  | isomer | H$_{vap}$ (eV) | dipole (D) | IE$_{vert}$ (eV) | β    |
|----|--------|----------------|------------|------------------|------|
| 10 |        | 0.342          | 19.0       | 1.56             | 1.21 |
|    | a      | 0.335          | 18.1       | 1.52             | 0.98 |
|    | b      | 0.323          | 2.5        | 1.40             | 2.00 |
| 12 |        | 0.331          | 21.1       | 1.47             | 1.16 |
|    | a      | 0.329          | 18.6       | 1.51             | 0.79 |
|    | b      | 0.322          | 19.2       | 1.49             | 0.91 |
|    | c      | 0.321          | 18.4       | 1.46             | 0.71 |
|    | d      | 0.318          | 18.0       | 1.47             | 0.68 |
|    | e      | 0.318          | 17.4       | 1.48             | 0.66 |
| 13 |        | 0.326          | 20.8       | 1.48             | 1.07 |
| 14 |        | 0.322          | 21.9       | 1.46             | 1.00 |
|    | a      | 0.318          | 15.0       | 1.50             | 0.77 |
|    | b      | 0.312          | 22.2       | 1.43             | 0.84 |
| 15 |        | 0.319          | 22.7       | 1.39             | 0.83 |
|    | a      | 0.318          | 21.1       | 1.45             | 0.90 |
|    | b      | 0.232          | 7.0        | 3.97             | 1.75 |
| 18 |        | 0.312          | 20.3       | 1.33             | 0.43 |
|    | a      | 0.308          | 18.5       | 1.38             | 0.38 |
| 20 |        | 0.318          | 21.1       | 1.35             | 0.52 |
|    | a      | 0.312          | 19.2       | 1.33             | 0.32 |

n=10: n=6 (octahedron) with four ligands added in the 2$^{nd}$ shell on one side (Li near the surface of the cluster). (a) similar to 10, but Li further away from the surface of the cluster. (b) n=6 (octahedron) with four ligands distributed symmetrically in the 2$^{nd}$ shell (Li at the center of the cluster).

n=12: n=6 (octahedron) with six ligands added in the 2$^{nd}$ shell on one side (Li near the surface of the cluster). (a) similar to n=12, but one ligand of the 2$^{nd}$ shell moved closer to the Li-side. (b) n=10 with one ligand added on the far side and one on the near side of Li. (c) similar to n=12b, but one ligand in the 2$^{nd}$ shell rearranged so that Li is more centered. (d, e) n=10b with two ligands added in the 2$^{nd}$ shell, distorting the symmetry.

n=14: n=6 (octahedron) with eight ligands added in the 2$^{nd}$ and 3$^{rd}$ shell on one side (Li near the surface of the cluster). (a) similar to n=12d, e with two more ligands added in the 2$^{nd}$ shell, Li slightly off-center. (b) n=14 with one ligand moved from the 2$^{nd}$ to the 3$^{rd}$ shell.

n=15: n=6 (octahedron) with nine ligands added in the $2^{nd}$ and $3^{rd}$ shell on one side (Li near the surface of the cluster). (a) similar to n=15, but ligands in the outer shell rearranged to yield an overall flatter cluster. (b) crystalline slab (monolayer) with Li at one edge.

n=18: n=6 (octahedron) with twelve ligands added in the $2^{nd}$ and $3^{rd}$ shell to yield a compact slightly flattened cluster shape with an off center Li closer to one surface. (a) similar to n=18, but more spherical in shape with Li closer to the center.

n=20: n=18 with two ligands added on the far side of Li. (a) n=18a with two ligands added, Li closer to the center.

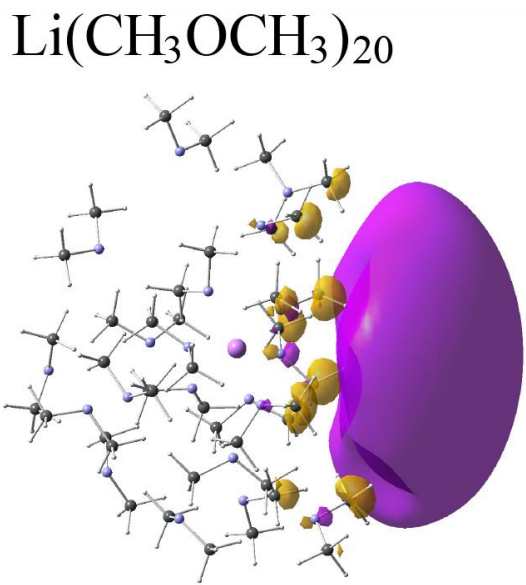

**Figure S1**: Isosurface of the HOMO of the most stable isomer of the Li(CH$_3$OCH$_3$)$_{20}$ cluster.